\documentstyle[pra,aps,epsfig]{revtex}
\begin{document}
\twocolumn[\hsize\textwidth\columnwidth\hsize\csname @twocolumnfalse\endcsname
\draft \author{E. Solano,$^{1,2}$ R. L. de Matos Filho,$^{1}$ and N. Zagury
  $^{1}$} \title{Mesoscopic superpositions of vibronic collective states of
  $N$ trapped ions} \address{$^{1}$Instituto de F\'{\i}sica, Universidade
  Federal do Rio de Janeiro, Caixa Postal
  68528, 21945-970 Rio de Janeiro, RJ, Brazil \\
  $^{2}$Secci\'{o}n F\'{\i}sica, Departamento de Ciencias, Pontificia
  Universidad Cat\'{o}lica del Per\'{u}, Apartado 1761, Lima, Peru}
\date{\today} \maketitle

\begin{abstract}
We propose a scalable procedure to generate superpositions of
motional coherent states and also entangled vibronic states in $N$
trapped ions. Beyond their fundamental importance, these states
may be of interest for quantum information processing and may be
used in experimental studies on decoherence.

\end{abstract}

\pacs{PACS number(s): 3.65.Ud, 42.50.Vk, 3.67.Hk }]

\vskip2pc

In quantum mechanics, the superposition principle, requiring that
all formal superpositions of quantum states give rise to new
physical states, plays a fundamental role~\cite{Dirac}. In
particular, it implies that probability densities of observable
quantities, in a superposition state, usually exhibit interference
effects instead of simply being added. This principle has been
successfully applied to explain a large class of new phenomena in
the microscopic level.  As an important example one can refer to
the prediction of superpositions of the $K^0$ meson and its
antiparticle as forming new and different states, the so called
$K^0_{1}$ and $K^0_2$ particles~\cite{Gell-Mann}.  Superpositions
of product states of composite systems also lead to the
fundamental quantum nonlocality, characteristic of quantum
entanglement. The quantum correlations resulting of such
superpositions lead to the violation of the Bell's
inequalities~\cite{Bell} and are essential ingredients for quantum
computation~\cite{qc} and quantum cryptography~\cite{crypto}.

Despite its experimental manifestations in the microscopic level,
the quantum superposition principle seems not to be applicable in
the macroscopic world, as it would predict, for example, the
existence of superpositions of a macroscopic object being at
different places. Such situation would stay in clear contradiction
to our observations and measurements of macroscopic phenomena. In
fact, the great majority of states predicted by the superposition
principle are not observed in our daily life. In recent years
decoherence\cite {Zurek} is being widely accepted as the mechanism
responsible for this fact and, consequently, for the emergency of
classicality. It consists in the fast decay of quantum
superpositions into statistical mixtures, and can be viewed as a
consequence of  always present interactions between a quantum
system and its environment. It precludes, in general, the
existence of macroscopic superpositions, except for very short
time intervals and is the key to understanding the fuzzy boundary
between quantum and classical behavior.  As coherence disappears
very fast as the system grows, it is important to be able to
generate mesoscopic superposition of states and analyze how it
looses its coherence. In this context, superpositions of coherent
states of both the electromagnetic field in a high~$Q$ microwave
cavity\cite{Haroche} and of the vibrational motion of a single
trapped ion\cite{Wineland2000} have been generated and their
behavior in the presence of the environment has been studied. Fast
generation of superposition of states involving a large number of
particles and excitations has become an experimental challenge in
order to understand how the transition from quantum to classical
occurs.

In this paper, we discuss some proposals for generating, in a fast
and controllable way, vibronic superpositions of mesoscopic states
involving, in principle, an arbitrary number $N$ of ions.  Relying
on homogeneous resonant bichromatic excitations of the $N$ ions
with laser light, we show that is possible to generate
superpositions of $N+1$ coherent states of the center of mass (CM)
vibration of the $N$ ions, equally spaced in phase space.
 We also show that it is possible to obtain mesoscopic
entangled states involving the internal and external degrees of
freedom of the $N$ ions and  odd and even coherent superpositions
of the CM motion of the ions, by additionally using a dispersive
bichromatic interaction. Our procedure opens the possibility of
generating mesoscopic superpositions of several massive particles
in a scalable and fast way.

Let us consider $N$ two-level ions of mass $m$, confined to move in the $z$
direction in a Paul trap. They are cooled down to very low temperatures~\cite
{king,sackett} and may perform small oscillations around their equilibrium
positions, $z_{j0},j=1,2...N.$ We denote by $Z=\sum_{j=1,N} z_j/N,$ the center
of mass coordinate and we set the origin at its equilibrium position. All ions
are simultaneously illuminated by two classical homogeneous Raman effective
pulses $\vec{E}_{I}= \vec{E}_{0I}e^{i(\vec{q}_{1}\cdot\vec{r}-\omega
  _{I}t-\varphi_I)}$ and $\vec{E}_{II} =
\vec{E}_{0II}\,e^{i(\vec{q}_{2}\cdot\vec{r}-\omega _{II}t-\varphi_{II})},$
with angular frequencies $\omega_{I}$ and $\omega_{II}$ and wave vectors
$\vec{q}_1=\vec{q}_{2}=\vec{q},$ parallel to the $z$ direction. The Raman
pulses frequencies will be chosen to be quasi-resonant with a long-living
electronic transition between two ionic hyperfine levels $|e_{j} \rangle $ and
$|g_{j} \rangle $ $(j=1,...N),$ with energies $\hbar\omega_0$ and $0$,
respectively. The Raman laser relative phases are chosen to be the same
$\varphi_I=\varphi_{II}=\varphi.$ The total Hamiltonian of the system may be
written, in the optical rotating wave approximation (RWA), as
\begin{equation}
\hat{H}=\hat{H}_0+\hat{H}_{{\rm int}},
\end{equation}
with
\begin{equation}\label{H0}
\hat{H}_0=\hbar \omega_0\sum_{j=1,N}|e_j \rangle \langle e_j| +
\hbar \nu \hat{a}^\dagger \hat{a} + \sum_{\lambda=1,N-1}
\hbar\nu_\lambda \hat{b}_\lambda^\dagger \hat{ b}_\lambda\, ,
\end{equation}
and
\begin{equation}\label{Hint}
\hat{H}_{\rm int}=\hbar \Omega \sum_{j=1,N}
e^{i(q\hat{z}_j{-\varphi})} | e_j\rangle\langle g_j|
\left(e^{-i\omega_{I}t}+ e^{-i\omega_{II}t}\right)+{\rm H.c.}
\end{equation}
The operators $\hat{a}$ and $\hat{b}_\lambda$ ($\hat{a}^{\dagger}$ and
$\hat{b}_\lambda^{\dagger }$) are the annihilation (creation) operators
associated with the center of mass mode of frequency $\nu$ and with the $N-1$
other vibrational modes of frequency $\nu_\lambda,$ respectively. For
simplicity, we have assumed that the same Rabi frequency $\Omega$ (taken as
real) is associated with both lasers.

We start by taking the frequencies $\omega_I$ and $\omega_{II}$ resonant with
the center of mass vibronic transition in the $k$-th blue and $k$-th red
sideband as
\begin{equation}  \label{freq}
\omega_{I} = \omega_0 + k\nu  \quad
{\rm and} \quad
\omega_{II} =\omega_0 -k\nu\, .
\end{equation}

For small $k$ values, we may safely assume that only the center of mass motion
will be excited, given that the next eigenfrequency is $\nu_r = \sqrt{3}\nu,$
corresponding to the stretch mode.  The following frequencies $(\ge\sqrt{29/5}
\nu) $depend on the number of ions and have being calculated in Ref.~
\cite{James}. Following the usual treatment for one single ion interacting
with a laser field~\cite{vogel}, we make the RWA with respect to the CM
vibrational frequency and select the terms that oscillate with minimum
frequency.  In the Lamb-Dicke limit the interaction
Hamiltonian may be written, in the interaction picture, as
\begin{equation}  \label{geral}
H_{\rm int}=\frac{2\hbar\Omega \eta^k}{k!}\hat{J}_{\rm T}(\hat{a}^k +
\hat{a}^{\dagger k })
\end{equation}
where $\eta =q\sqrt{\hbar /2Nm\nu }$ is the Lamb-Dicke parameter associated to
the center of mass motion of the $N$ ions, $\hat J_{\rm T}$ is an angular
momentum-like operator\cite{Nussenzveig} defined as
\begin{equation}\label{Jx}
\hat{J}_{\rm T}=\frac{i^k e^{-i\varphi}}{2}\sum_{j=1,N}\hat{\sigma}_{j+}
+{\rm H.c.}
\end{equation}
Here $\widehat{\sigma}_{+j}=|\uparrow_j\rangle\langle \downarrow_j |$
$=e^{iqz_{j0}} | e_j\rangle\langle g_j |$ is a flip operator associated with
the electronic transition $| g_j\rangle\rightarrow| e_j \rangle$ in the ion
$j$ and $\widehat{\sigma}_{j-}= \widehat{\sigma}_{j+}^\dagger.$ Without loss
of generality we may set the phase $\varphi=k\pi/2,$ so that $\hat{J}_{\rm
  T}=\hat{J}_x$ in the usual angular momentum operator convention for phases.
Similarly, by choosing a phase $\varphi=(k+1)\pi/2,$ $\hat{J}_{\rm
  T}=\hat{J}_y.$ We also may define the $z$ component of the angular momentum
by $\hat J_z=\frac12\sum_j (| e_j\rangle\langle e_j |-| g_j\rangle\langle g_j
|). $

From Eq.~(\ref{geral}), it is easy to show that the time evolution
operator, in the interaction picture, at time $t,$ is a sum of
products of unitary operators on the motional states and
projection operators on the ion internal states
\begin{equation}\label{evoltime}
\hat U_k(t)=\sum_{j,m }\hat  D_k(m\alpha_k(t))
|{j,m}\rangle_{x\, x}\langle{j,m }|
\end{equation}
where
\begin{equation}\label{displacement}
\hat{D}_k(\alpha_k)=e^{\alpha_k \hat{a}^{k\dagger} -\alpha_k* \hat{a}^k}\, ,
\end{equation}
with $\alpha_k(t)=2i\Omega t \eta^k/k!.$ Also, $|j,m\rangle_x$ are the
simultaneous eigenvectors of the operators $\hat J_x$ and $\hat J^2\equiv \hat
J_x^2+\hat J_y^2+\hat J_z^2,$ associated with the eigenvalues $m=-j,-(j-1),
.....j$ and $j(j+1),$ respectively.  $j$ varies from $0$ $(1/2)$ to $N/2$ by
steps of $1,$ if $N$ is even (odd).

The action of the time evolution operator $\hat U_k$ of
Eq.~(\ref{evoltime}) corresponds to unitary operations
$\hat{D}_k(m\alpha_k)$ on the motional degrees of freedom
conditioned to the value $m$ of the $x$ component of the "angular
momentum" electronic state. From now on we set $k=1$ that is, the
excitation occurs in the first red and blue sidebands. In this
case $\hat D_1(\alpha)$ is the displacement operator which
generates, when acting on the ground state,  coherent states of
the vibrational motion of the center of mass
\begin{equation}
|\alpha\rangle_{\rm coh}=e^{(\alpha \hat{a}^{\dagger} -\alpha* \hat{a})}
|0\rangle=e^{-|\alpha|^2/2}\sum \frac{\alpha^n}{\sqrt{n!}}|n\rangle\, .
\end{equation}

If initially the ions are in the vibronic ground state,
$|ggg...\rangle\otimes|0\rangle$ $\equiv |N/2,-N/2\rangle_z\otimes|0\rangle,$ their
state, after an interaction time $t$ with the laser fields, will be given by
\begin{equation}\label{Ncats}
\sum_{m,m'}d^{N/2}_{m',m}(\pi/2)
 d^{N/2}_{m,-N/2}(-\pi/2)
|{N/2,m'}\rangle_z\otimes|m\alpha\rangle_{\rm coh} ,
\end{equation}
where $\alpha=2i\eta\Omega t. $ Here, $d^{j}_{m',m}(\theta)=_z\langle j, m'|
e^{-i\theta J_y} |j, m\rangle_z$ are the matrix elements of the rotation
operator along the $y$ axis in the $\{|j, m\rangle_z\}$ basis of the
eigenstates of $\hat J_z$ and $\hat J^2$~\cite{Gottfried}. This allows us to
prepare in a very simple way and with a single Raman pulse, a mesoscopic
superposition of vibronic quantum states in $N$ trapped ions. If we now
measure the electronic state of the ions and find the totally excited state
$|eee...\rangle,$ we know that the motional state is given, up to a normalization factor,  by
\begin{equation}
\label{motionNcats}
\sum_{m=-N/2,N/2}\frac{(-1)^{(N/2-m)}}{(N/2-m)!(N/2+m)!}|m\alpha\rangle_{\rm
coh}.
\end{equation}

If $N$ is even, this state is a superposition of the vacuum and a series of
 coherent states of amplitudes $m\alpha,$ the probability for measuring
them decreasing with $|m|.$ If $N$ is odd, the vacuum is not
present in the superposition. When we have only one ion
Eq.~(\ref{motionNcats}) represents a single odd coherent state.
Note that the procedure above relies on a resonant excitation of
the ions and therefore can be accomplished with very {\it short}
interaction times.  The measurement of the electronic state can be
done by monitoring the fluorescence of a cyclic transition of the
ions~\cite{Dehmelt}, where a dark event detects the totally
excited state.  In Fig.~1, we show the Wigner function for the
state given by Eq.~(\ref{motionNcats}), when $|\alpha|=3$ and
$N=3.$ Note that the vacuum is not present, as   $N$ is odd, and
that interference manifests strongly, due to the multiple
superposition of coherent states.

\vspace*{-2.1cm}
\begin{figure}[htb]
\begin{center}
\hspace*{-1cm} \centerline{\leavevmode \epsfxsize=11cm
\epsfbox{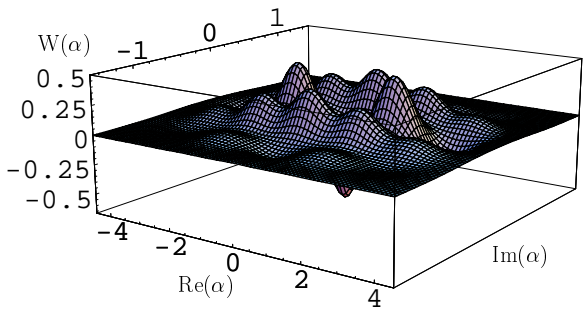}} \vspace*{-7cm} \caption{Wigner distribution for
the state given in Eq.~(\ref{motionNcats}) when $N=3$ and
$|\alpha|=3.$} \label{fig1}
\end{center}
\end{figure}

If the frequencies of the lasers in Eq.~(\ref{Hint}) are slightly off
resonance and such that $\omega_I=\omega_0 +\nu+\delta$ and
$\omega_{II}=\omega_0 -\nu-\delta,$ the time evolution operator may be
written, for small values of $\eta,$ approximately as $e^{-i\lambda t
  \hat{J}_y^2},$ where $\lambda= 8( \Omega\eta)^2 /\delta $ and we have chosen
$\varphi=0.$ This dispersive bichromatic interaction does not change the
motional state and has been studied in a series of recent
papers\cite{sorensenbell,sorensenGHZ,sorensenres,SMZ}. It has been implemented
successfully in the laboratory for generating GHZ states in two and four
ions\cite{sackett}.

We now show that, by using both the dispersive and the resonant bichromatic
interaction, it is possible to prepare even and odd coherent states of large
amplitudes and involving a large number of ions.  It will be convenient to
consider first the case where we have an odd number of ions $N.$ If we start
with the ground state $|ggg...\rangle\otimes|0\rangle$ and apply the
dispersive bichromatic interaction during a time $t=\pi/(2\lambda),$ we get,
using the properties of the rotation matrices\cite{Gottfried},
\begin{equation}\label{Nodd}
e^{-i\hat{J}_y^2\pi/2}|j,-j\rangle_z\otimes|0\rangle=
\frac{1}{\sqrt{2}}(|j,j\rangle_x-|j,-j\rangle_x)\otimes|0\rangle
\, ,
\end{equation}
where $j=N/2.$

Now we apply a second pulse using the resonant lasers of frequencies
$\omega_0\pm\nu,$ with a relative phase of $\pi/2$ with respect to the
previous pulse, during a time $\tau.$ This corresponds to apply the operator
$U_1(\tau)$ on the state given in Eq. \ref{Nodd} leading to the state
\begin{eqnarray}\label{firstcat}
&\frac{1}{\sqrt 2}&(|N\alpha/2\rangle_{\rm coh}\otimes|N/2,N/2\rangle_x \nonumber\\
&-&|-N\alpha/2\rangle_{\rm coh}\otimes|
|N/2,-N/2\rangle_x \, .
\end{eqnarray}
where $\alpha=2i\Omega\eta\tau.$ This state is a superposition of
two vibronic states involving, in principle, a large number $N$ of
ions. The two internal states $|N/2,\pm N/2\rangle_x$  are
strongly correlated to the two vibrational states $|\pm
N\alpha/2\rangle_{\rm coh},$ whose average amplitude of
oscillation, $|N\alpha/2 |,$ is proportional both to $\sqrt{N}$
and $\Omega\tau.$ This state is an example of a strongly entangled
mesoscopic state that is scalable, and its experimental
realization may become a useful tool in analyzing the dependence
of decoherence on the number of degrees of freedom of this system.

If, after the preparation of the state given in Eq.~(\ref{firstcat}), we
measure the state $|eee...\rangle$, we obtain
\begin{equation}\label{secondcat}
\frac{|N\alpha/2\rangle_{\rm coh}
+|-N\alpha/2\rangle_{\rm coh}}{\sqrt{2+2e^{-|N\alpha|^2/2}}} \, .
\end{equation}
The state given in Eq.~(\ref{secondcat}) is an even coherent state of the
center of mass motion of an odd large number of ions.  An odd coherent state
is obtained if one measures the state $|ggg...\rangle$ instead of
$|eee...\rangle.$ Thus, we have a procedure that generates a superposition of
two states of motion, vibrating out of phase, where a large number of
particles may be involved.  In principle, the number of ions in this state is
limited mostly by decoherence and by the capacity of producing lasers with
high homogeneity. For large $N,$ the probability to produce these state
superpositions by this procedure decreases as ${1/2}^N.$ However, an efficient
method to obtain even or odd coherent states, each one with probability $1/2,$
even for large $N,$ will be described below.

As before, we start with the ground state and apply in succession the
dispersive bichromatic interaction as in Eq.~(\ref{Nodd}) and the resonant
bichromatic interaction to obtain the state given in Eq.~(\ref{firstcat}). We
then apply again the same dispersive bichromatic interaction during the same
interval of time. The resulting state is

\begin{eqnarray}
&&\frac{1}{2}\{|eee..\rangle\otimes(|N\alpha /2\rangle
-|-N\alpha /2\rangle ) \nonumber \\
&&-|ggg...\rangle\otimes(|N\alpha /2\rangle +|-N\alpha /2\rangle )\}
\end{eqnarray}
Measuring the dark fluorescence of either the $N-$ electronic excited state or
the $N-$ electronic ground state, we get the associated odd or even coherent
state with probability 1/2. This procedure shows how to generate efficiently
mesoscopic superposition states for a large number of ions.

Similar results may be obtained for $N$ even, if an additional
resonant carrier $\pi/2$ pulse, is applied at the same time and
with a relative phase $\varphi=\pi/2$ with respect to the
dispersive pulse.

In conclusion, we have presented a procedure to generate several
kinds of mesoscopic superpositions of states involving $N$ ions.
Superpositions of coherent states evenly spaced on a line may be
generated rapidly, through resonant interactions.  Even or odd
coherent state may be obtained if we use also dispersive
bichromatic interactions.  This is, to  the best of our knowledge,
the first proposal to generate, in a  scalable way, mesoscopic
superpositions of collective motional  states of $N$ trapped ions.
Note that,  as long as  the field is spatially homogeneous over
the trapped ions, the only parameter that depends on $N,$ in our
model,  is the time we applied the laser pulses, which varies with
$\sqrt{N}(N)$ during the resonant (dispersive) interaction. We
also expect that the cooling of the vibrational modes for $N$ ions
will be achieved with the same effectiveness as in the case of
four ions\cite{sackett}.  For this reason the constraints, in our
proposal, on parameters like laser pulse timing, laser
frequencies, as well as cooling to the ground state should remain
almost the same as in the case of a single trapped ion. Then, the
number of ions involved will be limited mostly by the scale of
time in which decoherence occurs.

We believe that the proposals presented in this letter are ready
to be implemented in the laboratory. They should help to build
larger mesoscopic quantum superpositions in trapped ions, to study
experimentally, at large scale, decoherence processes and for
applications in quantum information processing.

This work was partially supported by the Conselho Nacional de Desenvolvimento
Cient{\'\i}fico e Tecnol\'ogico (CNPq), the Funda\c{c}\~ao de Amparo a
Pesquisa do Estado do Rio de Janeiro (FAPERJ), the Programa de Apoio a
N\'ucleos de Excel\^encia (PRONEX) and Funda\c{c}\~ao Jos\'e Bonif\'acio
(FUJB).

\end{document}